# Paraboloidal Crystals

Luca Giomi, Mark Bowick
*Syracuse University, Syracuse, NY 13244*

The interplay between order and geometry in soft condensed matter systems is an active field with many striking results and even more open problems. Ordered structures on curved surfaces appear in multi-electron helium bubbles, viral and bacteriophage protein capsids, colloidal self-assembly at interfaces and in physical membranes[1].

Spatial curvature can lead to novel ground state configurations featuring arrays of topological defects that would be excited states in planar systems[2].

The sequence of images shows the Voronoi lattice (in gold) and the Delaunay triangulations (in green) for ten low energy configurations of a system of $N \in [16, 200]$ classical charges constrained to lie on the surface of a paraboloid and interacting with a Coulomb potential. The parabolic geometry is considered as a specific realization of the class of crystalline structures on two-dimensional Riemannian manifolds with variable Gaussian curvature and boundary.

The coexistence of isolated disclinations (7-fold in blue, 5-fold in red) and grain boundary "scars" (arrays of tightly-bound (5,7)-fold disclination pairs) illustrates the rich structure arising from the novel geometry and topology of the paraboloid[3].

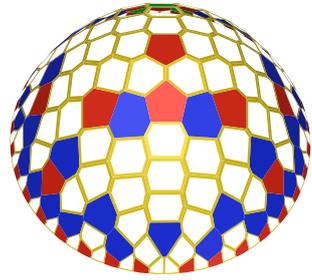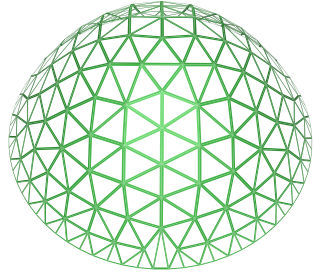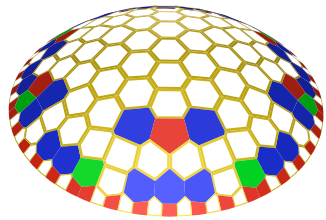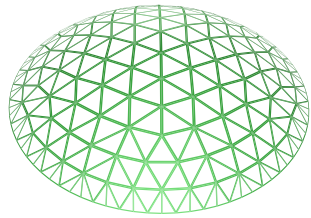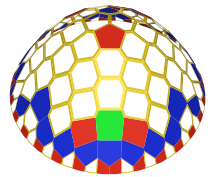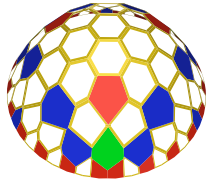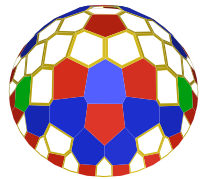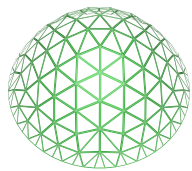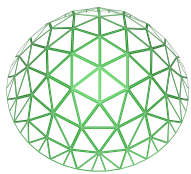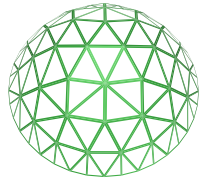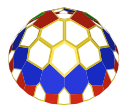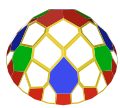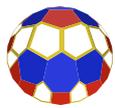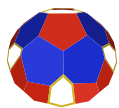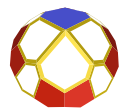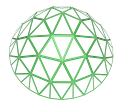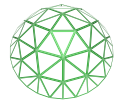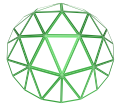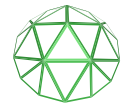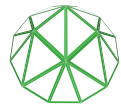